%% file: imp.tex
\documentstyle[psfig]{europhys} 

\newcommand{\lsim}{\mbox{\raisebox{-0.1cm}{$\;
\stackrel{\textstyle<}{\sim}\;$}}}
\input euromacr

\begin{document}
\euro{xx}{x}{xx-xx}{xxxx}
\Date{}
\shorttitle{E. CAPPELLUTI {\it et al.} IMPURITY EFFECTS ON THE FLUX PHASE ETC.}
\title{{\Large Impurity Effects on the Flux 
Phase Quantum Critical Point Scenario}} 
\author{E. Cappelluti$^*$ and
R. Zeyher}
\institute{ Max-Planck-Institut f\"ur Festk\"orperforschung,
Heisenbergstr. 1, 70569 Stuttgart, Germany}
\rec{}{}
\pacs{
\Pacs{74}{10$+v$}{Occurrence, potential candidates}
\Pacs{74}{25$Dw$}{Superconductivity phase diagrams}
\Pacs{74}{20$Mn$}{Nonconventional mechanisms}
}

\maketitle
\begin{abstract}
Impurity substitution of Zn in La-214 and (Y,Ca)-123 high-$T_c$ 
superconductors suppresses $T_c$ but does not affect appreciably
the onset of the pseudogap phase in the underdoped region nor
optimal doping or the position of the inferred quantum critical point.
Based on a $1/N$ expansion of the $t-J$ model we explain these findings 
as well as the similar dependence on a
magnetic field in terms of a quantum critical point scenario where a flux 
phase causes the pseudogap.
\end{abstract}

The quantum critical point scenario\cite{monthoux,castellani,varma}
 represents a popular frame to discuss
the phase diagram of high-$T_c$ oxides. Suppressing superconductivity by
strong magnet fields it has been found experimentally\cite{boebinger}
 that there
exists a critical hole doping $\delta^{QCP}$ at zero temperature which
separates a metallic state at larger dopings from an insulating state
at lower dopings. Strong fluctuations of the order parameter related to
the insulating phase are thought to suppress the density of states
for $\delta < \delta^{QCP}$ leading to the pseudogap features in the underdoped
region and to be instrumental for superconductivity around $\delta_c$
and, at higher temperatures, for the anomalous properties of the normal
state in these systems.

The microscopic nature of the order parameter of the insulating phase
and its fluctuations are presently not clear. One obvious choice is 
antiferro\-magnetism\cite{monthoux} which occurs at $T=0$ as long-range 
ordered  phase
at zero and small dopings. The corresponding zero temperature critical point,
however, corresponds to a much smaller doping value than the observed
one, $\delta^{QCP} \sim 0.17$. A reasonable
large $\delta^{QCP}$ has been obtained in Ref. \cite{castellani}
for a scenario with an
incommensurate charge density wave (ICDW).
In this approach the pseudogap features are not directly related to 
the  ICDW order parameter but rather connected to strong $d$-wave
superconducting fluctuations sustained by ICDW precursors. 
Related approaches include preformed Cooper pairs
where phase coherence is achieved below
$T_c$\cite{emery} or RVB spinon pairing\cite{fukuyama} and the $\pi$-flux
phase\cite{wen}, where charge coherence is obtained by Bose condensation
of holons. A different proposal has been made in Ref. \cite{czprb}.
Based on a $1/N$ expansion for
the $t-J$ model the quantum critical point was identified with a transition
from the normal to a $d$-wave flux state occurring near the observed 
$\delta^{QCP}$
for realistic parameters. In this approach optimal doping is determined
by the onset of the flux phase and the phase diagram in the underdoped 
region is characterized by the competition between the flux and 
superconducting order parameters both having $d$-wave symmetry. 

Recently, several experimental results have been published which may
be able to confirm or to rule out some of the above approaches. Measurements
of NMR spin lattice relaxation rates in the presence of magnetic fields 
up to $\sim 15$ Tesla did not yield appreciable changes for the onset
temperature $T^*$ of the
pseudogap phase whereas the superconducting $T_c$ was reduced by 
about $8$ K \cite{gorny}. 
A strong suppression of $T_c$ and, at the same time, no change in the
pseudogap was previously reported in Zn-doped
YBa$_2$Cu$_3$O$_{6+x}$\cite{alloul}.
High resolution photoemission\cite{ino}, electronic 
Raman spectroscopy\cite{naeini}, NMR\cite{williams}
and heat capacity data\cite{loram,williams2} show that $T^*$ does not merge
with $T_c$ in the overdoped regime, but vanishes near 
optimal doping. These findings indicate that the pseudogap and the 
superconductivity are different phenomena and not related to the
same order parameter. Furthermore, it has been found 
experimentally\cite{tallon}
that the lowering of the $T_c$ curves in Zn doped 
Y$_{0.8}$Ca$_{0.2}$Ba$_2$Cu$_3$O$_7$ ((Y,Ca)-123) 
and La$_{2-x}$Sr$_x$CuO$_4$ (La-214), 
is concentrated around optimal doping and that the optimal doping itself
is not shifted. 
It is the purpose of this Letter to investigate the
influence of impurity scattering and magnetic fields on the phase diagram
calculated in Ref. \cite{czprb} and to compare the results with the above experimental
findings.

\begin{figure}[t]
\protect
\centerline{\psfig{figure=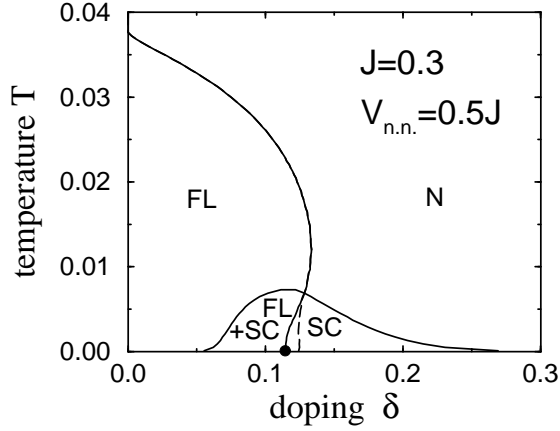,width=8cm}}
\caption{Phase diagram of the $t$-$J$-$V$ model (see text)
taking into account the normal (N), superconducting (SC), and the 
flux phase (FL) as well as the coexisting state of superconductivity
and flux phase (FL+SC). The normal state-flux phase transition
ends in a $T = 0$ quantum critical point at $\delta^{QCP} \simeq 0.115$.
The dashed line represents the instability towards the flux phase
in the superconducting state.}
\label{fig1}
\end{figure}

We consider a $t$-$J$-$V$ model with $N$ degrees of freedom per lattice
site on a square lattice. Its Hamiltonian can be written in terms of 
Hubbard's $X$-operators as \cite{czprb}
\begin{eqnarray}
H & = & - \frac{t}{N} \sum_{\langle i j \rangle \atop p=1 \ldots N}
X_{i}^{p 0} X_{j}^{0 p} +\frac{J}{4N}
\sum_{\langle i j \rangle \atop p,q =1 \ldots N}
 X_{i}^{p q} X_{j}^{q p}\nonumber\\
 & & - \frac{J}{4N}\sum_{\langle i j \rangle \atop p,q =1 \ldots N}
 X_{i}^{p p} X_{j}^{q q}+
\sum_{i j \atop p,q =1 \ldots N}
\frac{V_{i j}}{2 N} X_{i}^{p p} X_{j}^{q q}\:.
\label{htj}
\end{eqnarray}
The internal labels $p$,$q$... consist of a spin label distinguishing
spin up and spin down states and a flavor label counting $N/2$ identical
copies of the original orbital. $\langle i j \rangle$ denotes 
nearest-neighbor sites. 
The first three terms represent the $t$-$J$ Hamiltonian,
the last term a screened Coulomb interaction appropriate for two
dimensions and taken from Ref. \cite{becca}.
In the following we express all energies in units of $t$.
The strength of the Coulomb interaction will be characterized by its
value between nearest neighbor sites $V_{n.n.}$.
In the limit of large $N$, the interactions become purely
instantaneous and $H$ can be diagonalized analytically.
In the absence of impurity scattering,
the coexistence state of superconductivity and a staggered $(\pi,\pi)$
flux phase can be obtained from a Nambu representation with 4 states
yielding four electronic bands with dispersion \cite{czprb}
\begin{equation}
\pm E_{\pm}({\bf k}) = \pm \sqrt{[\xi({\bf k}) 
\pm \tilde{\mu} ]^{2}+\Delta({\bf k})^{2}}\:,
\end{equation}
where
\begin{equation}
\xi({\bf k}) = \sqrt{\epsilon({\bf k})^{2}+
\phi({\bf k})^{2}}\:.
\end{equation}
Here the momenta $\bf k$ are restricted to the new Brillouin zone
which is one half of the original one. $\tilde{\mu}$ is a 
renormalized chemical potential, $\phi({\bf k})$ the 
flux order 
parameter, $\Delta({\bf k})$ the superconducting gap, and
$\epsilon({\bf k})$ the one-particle energies in the normal state.
Both order parameters have $d$-wave symmetry:
$\phi({\bf k}) = \phi [\cos(k_x)-\cos(k_y)]$, 
$\Delta({\bf k}) = \Delta [\cos(k_x)-\cos(k_y)]$.
They are determined by the self-consistent set of equations:
\begin{equation}
\phi({\bf k}) =  \frac{1}{2 N_{c}} \sum_{{\bf p}}
J({\bf k+p})\, \eta_{\phi}({\bf p})\:,
\label{phi}
\end{equation}
\begin{equation}
\Delta({\bf k}) = \frac{1}{2 N_{c}} \sum_{{\bf p}}
[ J({\bf k+p}) - V_{n.n.}({\bf k+p})]\, \eta_{\Delta}({\bf p})\:,
\label{delta}
\end{equation}
where
\begin{equation}
\eta_{\phi}({\bf k}) =
\frac{\phi({\bf k})}{\xi({\bf k})}
\left\{
\frac{\xi({\bf k})+\tilde{\mu}}{2E_{+}({\bf k})}
\tanh\left[\frac{E_{+}({\bf k})}{2 T}\right]
+ \frac{\xi({\bf k})-\tilde{\mu}}{2E_{-}({\bf k})}
\tanh\left[\frac{E_{-}({\bf k})}{2 T}\right]\right\},
\label{etaphi}
\end{equation}
\begin{equation}
\eta_{\Delta}({\bf k}) =
\frac{\Delta({\bf k})}{2E_{+}({\bf k})}
\tanh\left[\frac{E_{+}({\bf k})}{2 T}\right]
+\frac{\Delta({\bf k})}{2E_{-}({\bf k})}
\tanh\left[\frac{E_{-}({\bf k})}{2 T}\right].
\label{etadelta}
\end{equation}
The resulting phase diagram, calculated using $J=0.3$ and $V_{n.n.}= 0.5J$,
is shown
in Fig. \ref{fig1}. Disregarding superconductivity, the second-order 
normal state-flux phase transition line ends in 
a quantum critical point, denoted by
the black dot, at $\delta^{QCP} \sim 0.115$.
We find the maximum of $T_c$
at essentially the same doping because of the strong competition between
flux and superconducting phase\cite{czprb},
and also that
the flux phase instability is only slightly shifted by
superconductivity (dashed line).

Now we are going to investigate how the phase diagram
in Fig. \ref{fig1} is affected by impurity scattering.
In the simplest approximation, the effects of impurities 
in the normal state can be
taken into account by introducing a renormalized frequency\cite{abrikosov}
\begin{equation}
i\tilde{\omega}_n = i\omega_n  + i \Gamma \frac{\omega_n}{|\omega_n|},
\label{imp1}
\end{equation}
where $\Gamma$ is a scattering rate, here used as a
free parameter proportional to the impurity concentration.
Throughout the flux phase the self-energy due to impurity scattering
is still diagonal in the 4x4 Nambu representation because the
flux order parameter does not couple to the impurities. 
The constant $\Gamma$ in Eq. (\ref{imp1}) could be improved 
in a deeper analysis considering effects due to the proximity of the Van Hove
singularity\cite{fehrenbacher}. 
However, the interesting doping region for superconductivity is in our model
not at all correlated with the Van Hove
singularity. As a matter of fact, the chemical potential for 
$\delta \sim 0.1$ is quite far away from the Van Hove singularity, 
which is at $\delta = 0$ in our model.
In this situation the band can be assumed in a good approximation 
to be structureless and $\Gamma$ as a constant,
even in the case of strong potential scattering\cite{fehrenbacher}.
Experimentally it is known that non-magnetic impurities such as Zn and Al
induce local moments on neighboring Cu sites\cite{alloul,ishida}.
Both in a $d$-wave flux phase and a $d$-wave superconductor random local
magnetic moments lead only to renormalizations of the frequency. As a
result they contribute additively to $\Gamma$ in Eq.(\ref{imp1}) 
and thus can be accounted for by a proper choice for $\Gamma$. 
It also has been argued\cite{ishida} that strong potential scattering
near the unitary limit is much more important for the reduction of $T_c$
than the scattering from induced magnetic moments.

The occupation number $f_\Gamma$ of an electronic state with energy
$\epsilon$ in the presence of impurities reads 
\begin{eqnarray}
f_\Gamma\left(\frac{\epsilon}{T}\right) &=& 
T \sum_{n} \frac{e^{i\omega_n 0^+}}{i\tilde{\omega}_n - \epsilon} 
\nonumber\\
&=&
\frac{1}{2} 
+\frac{1}{2\pi} \mbox{Im}\left[
\psi\left(\frac{1}{2}-i\frac{\epsilon + i \Gamma}{2\pi T}\right)\right.
- \left.\psi\left(\frac{1}{2}+i\frac{\epsilon - i \Gamma}{2\pi T}\right)
\right],
\label{fgamma}
\end{eqnarray}
where $\psi$ denotes the digamma function.
In the limit of zero impurity concentration $\Gamma \rightarrow 0$ and
$f_\Gamma(\epsilon /T)$ reduces to the usual Fermi function $f(x)=1/(e^x+1)$.
In a similar way, we also define a function $\tanh_\Gamma$ by
\begin{equation}
\tanh_\Gamma\left(\frac{\epsilon}{2T}\right)=
\frac{1}{2}\left[ f_\Gamma\left(-\frac{\epsilon}{T}\right)-
f_\Gamma\left(\frac{\epsilon}{T}\right)\right].
\label{tanhgamma}
\end{equation}

\begin{figure}[t]
\protect
\centerline{\psfig{figure=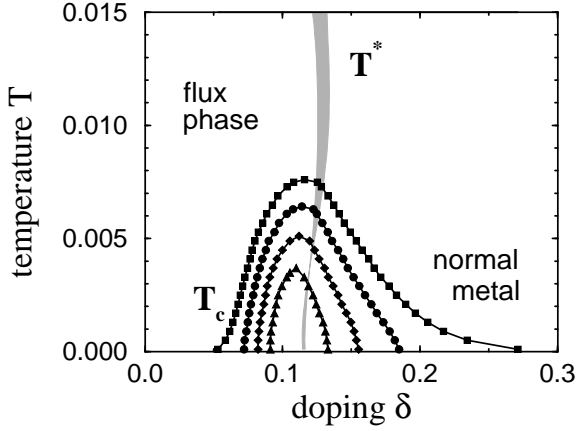,width=8cm}}
\caption{Solid lines: Suppression of $T_c$ by impurity scattering, 
calculated with the scattering rates $\Gamma$ =
0 (squares), $2\cdot 10^{-3}$ (circles), 
$4\cdot 10^{-3}$ (diamonds), and 
$6\cdot 10^{-3}$ (triangles). Grey region: Corresponding variation of the
transition temperature $T^*$ to the flux state.}
\label{fig2}
\end{figure}

For the determination of the superconducting critical temperature
$T_c$, the self-consistent set of gap equations 
can be linearized with respect to $\Delta({\bf k})$. The resulting equations
are again given by Eqs. (\ref{phi}-\ref{etadelta}) if the function 
$\tanh$ is everywhere replaced by the function   
$\tanh_\Gamma$, defined in Eq. (\ref{tanhgamma}). The
solid lines in Fig. \ref{fig2} show numerical results for $T_c$ as a function
of doping $\delta$ for different scattering rates $\Gamma$, using $J=0.3$
and $V_{n.n.}=0.5J$. These curves
illustrate the suppression of $T_c$ with increasing scattering rates
$\Gamma = 0$, $2\cdot 10^{-3}$, $4\cdot 10^{-3}$, 
and $6\cdot 10^{-3}$. The corresponding changes in $T^*$, determining the 
phase boundary between the normal state and the flux state, are depicted in 
Fig. \ref{fig2} by the grey region. The chosen values for 
$\Gamma$ correspond roughly to 
$\Gamma \simeq 1.0 T_c$ at optimal doping, and
to $\Gamma \simeq 1.5 T_c$ in the strongly underdoped region, interpolating
between the weak- and the strong-coupling regimes.
One important result of Fig. \ref{fig2} is that the flux phase boundary
$\delta^{FL}(T)$ is only slightly shifted by impurities, in spite of
the strong suppression of the superconducting critical temperature.
In particular, the zero temperature limit of $\delta^{FL}$, $\delta^{FL}(0)$,
is almost completely independent of the impurity scattering rate. 
Since in our approach the maximum of $T_c$ as a function of doping 
is essentially determined by $\delta^{FL}(0)$ this means that 
the $T_c(\delta)$ curves shrink to $\delta^{FL}(0)$ with increasing
scattering rate which is a characteristic feature of Fig. \ref{fig2}.
Interpreting Fig. \ref{fig2} in terms of a quantum critical point scenario
means that the corresponding critical doping $\delta^{QCP}$ is given by
$\delta^{FL}(0)$ and that $\delta^{QCP}$ is almost completely independent 
of the 
impurity scattering rate. The curves in Fig. \ref{fig2} are in excellent
agreement with the corresponding experimental curves in Zn doped 
(Y,Ca)-123 and La-214,
given in Fig. 2 of Ref. \cite{tallon}. 

We can gain further insight into our results by the following analysis.
We first notice
that, at least for the above values of $\Gamma$,
impurity scattering effects lead to an additional
smearing of the occupation number 
in Eq. (\ref{fgamma}) which can be simulated in a very good approximation 
by an effective temperature:
\begin{equation}
f_\Gamma\left(\frac{\epsilon}{T}\right) \simeq
f\left(\frac{\epsilon}{T+\Gamma}\right).
\end{equation}
As a consequence, the flux phase instability in the presence of impurities
is roughly determined by
$\delta^{FL}_{\Gamma}(T) \simeq \delta^{FL}(T+\Gamma)$.
But $\delta^{FL}$ is very weakly dependent on $T$ until $T \sim \tilde{\mu}$,
so that the $T = 0$ quantum critical point is not expected to be
shifted as long as $\Gamma \lsim \tilde{\mu}$ holds.

\begin{figure}[t]
\protect
\centerline{\psfig{figure=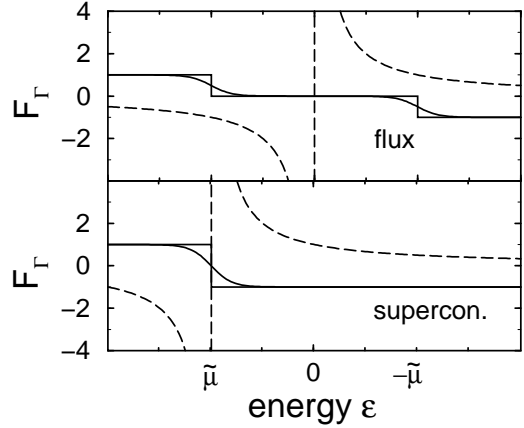,width=8cm}}
\caption{Numerator (solid lines) and denominator (dashed line)
of $F_\Gamma(\epsilon)$, Eq. (\ref{fflux}),
in the case of the flux instability of the normal state (upper panel).
The discontinuous and continuous solid lines refer to 
zero and nonzero effective temperature, respectively.
For comparison, the lower panel shows the corresponding quantities
in the case of a superconducting instability.}
\label{fig3}
\end{figure}

In order to understand better why $\delta^{FL}(T)$ is almost 
independent of $T$ for $\Gamma \lsim \tilde{\mu}$, we consider the
flux phase susceptibility $\chi(T,\Gamma,\delta)$.
The instability line $\delta^{FL}(T)$ is determined 
in general by the equation:
$1 = J \chi(T,\Gamma,\delta)$,
where the dependence of the susceptibility on temperature and impurities is
given by the factor\cite{czprb}
\begin{equation}
F_\Gamma(\epsilon) = \frac{f_\Gamma[(\epsilon-\tilde{\mu})/T]-
f_\Gamma[(-\epsilon-\tilde{\mu})/T]}{\epsilon}.
\label{fflux}
\end{equation}
A sketch of the numerator and denominator of $F_\Gamma$ is given in 
Fig. \ref{fig3}.
Continuous and discontinuous solid lines represent the numerator for 
$\Gamma \neq 0$ and $\Gamma = 0$, respectively, the dashed line the 
denominator.
Due to the weak variation of the denominator around $\tilde{\mu}$ and 
$-\tilde{\mu}$, it is 
clear that this factor is only slightly affected by a possible smearing
due to finite temperatures or impurity concentrations,
as long as $T+\Gamma \lsim \tilde{\mu}$ holds. Things are different 
in the case of the superconducting
susceptibility. Here  the divergence of the denominator coincides with the
jump in the numerator, so that even a small smearing 
leads to a strong change in $\chi$ and a large suppression of $T_c$.

We would like to mention that a charge-density-wave susceptibility
would contain a similar factor as in Eq. (\ref{fflux}), so that
also in this case impurities will not affect substantially the
function $F_\Gamma$. However, the
CDW order parameter has the symmetry of the underlying
lattice, i.e., $s$-wave symmetry. Impurities couple in this case directly
to the order parameter and the self-energy due to impurity scattering
also acquires non-diagonal elements in addition to the diagonal ones
described by Eq. (\ref{imp1}). As a result, one expects that the charge-density
wave state is sensitive to impurities and the
corresponding quantum critical point and optimal doping would be shifted
by the impurities, in disagreement with Ref. \cite{tallon}. 

We have considered throughout our analysis a $(\pi,\pi)$ flux phase
and its competition with superconductivity. In Ref. \cite{czpc}
it was shown that there is a continuous transition line in the $T-\delta$ plane
from a commensurate to an incommensurate flux state at low temperatures
(the term ``commensurate'' is here used with respect to the lattice
periodicity and not, as in Ref. \cite{lederer}, with respect to the
electronic filling).
Taking the incommensurability into account 
the largest onset of the
flux phase as a function of doping occurs now at $T=0$ with a critical
doping $\delta^{QCP} \sim 0.135$. 
However, the boundaries
in the phase diagram and, in particular, the competition between flux and 
superconducting phases are not much changed by allowing for the 
incommensurability of the flux phase. Disregarding superconductivity 
we also have studied the influence of impurities on the boundary between
an incommensurate flux and the normal phase. The resulting width in $T^*$
for the scattering rates used in Fig. 2 is very similar to that shown in
this Figure for the commensurate case. In particular, the change of the
critical doping at $T=0$ was for all $\Gamma$'s smaller than $0.01$
showing that our Fig. 2 calculated for the commensurate case
is also valid in the incommensurate case in a very good approximation.
Also the simplified arguments based on Fig. 3 for the
robustness of the flux state in contrast to the superconducting state
with respect to impurities still apply. The finiteness of the chemical
potential $\tilde{\mu}$ reflects the fact that one-particle
states which are also not exactly degenerate in energy are involved in
forming the flux state. This means that finite 
difference energies $|\epsilon({\bf k})-\epsilon({\bf k}+{\bf Q})|$
($\bf Q$ is the wave vector of the flux phase) 
associated with a large phase space are important which can be characterized
by a typical energy $2 \tilde{\mu}$. This explains why both
the commensurate and the incommensurate flux phase behave in a very similar
way with respect to impurities.

Our proposed scenario of a flux quantum critical point is also
consistent with the NMR measurements of Ref. \cite{gorny}.
In that paper, a magnetic field $H = 14.8$ T was shown to yield 
a net reduction of the superconducting critical temperature 
of $\Delta T_c = 7.8$ K but no corresponding decrease 
of the pseudogap temperature $T^*$ within the experimental 
uncertainty of 2 $\%$.
In our theory the pseudogap and the superconductivity arise from
two different mechanisms, so that a significant reduction
of $T_c$ is possible in the absence of a corresponding reduction
of $T^*$. The predominant effect of a magnetic field 
on the superconducting phase is a reduction of $T_c$ 
in order to balance the free magnetic energy related to the Meissner effect.
The decrease of $T_c$ is linear in $H$ for $H \ll H_c$
($H_c$ being the critical magnetic field),
so that the reduction in $T_c$ is quite effective even for small magnetic 
fields.
On the other hand, the effect of a magnetic field on the flux phase
is mainly due to the Zeeman splitting $\Delta E = g \mu_B H$, 
where $g$ and $\mu_B$ are the $g$ factor and the Bohr magneton, respectively.
This energy is about $20$ K for $H = 14.8$ T\cite{gorny}, and thus
much smaller than the width of the electronic band.
If we generalize our order parameter $\phi$  in the presence of a
magnetic field via $\phi \rightarrow \phi = 
[\phi_\uparrow(\tilde{\mu}-\Delta E)+\phi_\downarrow(\tilde{\mu}+\Delta E)]/2$,
we obtain an effective susceptibility
given by $\chi =
[\chi_\uparrow(\tilde{\mu}-\Delta E)+\chi_\downarrow(\tilde{\mu}+\Delta E)]/2$.
The Zeeman splitting $\Delta E = 20$ K is much smaller than the
energy scale set by the bandwidth $W \approx 0.5$ eV, so that we expect
a negligible effect on the flux instability.
Moreover the shift of $T_c$ will be only of order $H^2$ in the magnetic field.

We have checked the above arguments by calculating explicitely
the chan\-ge in the instability from the normal to the flux state in the 
presence of a Zeeman splitting
$\Delta E = 20$ K, corresponding to $\Delta E = 6 \cdot 10^{-3} t$ with
$t = 0.3$ eV. We find a shift in doping of the
$T = 0$ quantum critical point of about 4 $\%$, which is quite small.
This shift disappears with increasing temperature 
because of thermal smearing which makes the Zeeman splitting ineffective.

In conclusion, we have analyzed the effects of impurity scattering on the flux
phase and on its interplay with superconductivity within
the framework of a quantum critical point scenario.
We have found that the transition between the flux and the normal phase
is essentially unaffected by impurities and magnetic fields, in very
good agreement with the experimental data. This is especially true
at zero temperature where we identify the quantum critical point with
the transition between the normal and the (incommensurable) flux state.
We also pointed out that a charge density wave as the origin of the 
pseudogap phase would directly couple to impurities in contrast to the
flux order parameter and thus be much more sensitive to impurities.

\stars{}

\vskip12pt

$^*$ Present address: Dipartimento di Fisica, Universit\`a di Roma I
``La Sapienza'', P.le Aldo Moro 2, 00184 Roma, Italy.

\end{document}
\bye

%% file: euromacr.tex
\def\And{{\rm and\ }}

\def\stars{\bigskip\centerline{***}\medskip}

\newif\ifboo \boofalse

\def\Review#1{\boofalse{\it #1},}
\def\Name#1{{\sc #1},}
\def\Vol#1{\ifboo Vol. {\bf #1}\else{\bf #1}\fi}
\def\Year#1{\ifboo #1\else(#1)\fi}

\def\Page#1{\ifboo {\rm p. #1}\else{\rm #1}\fi}